\documentstyle[12pt]{article}
\def\PRL #1 #2 #3{{\sl Phys. Rev. Lett.} {\bf#1} (#2) #3}
\def\NPB #1 #2 #3{{\sl Nucl. Phys.} {\bf B #1} (#2) #3}
\def\NPBFS #1 #2 #3 #4{{\sl Nucl. Phys.} {\bf B #2} [FS#1] (#3) #4}
\def\CMP #1 #2 #3{{\sl Commun. Math. Phys.} {\bf #1} (#2) #3}
\def\PRD #1 #2 #3{{\sl Phys. Rev.} {\bf D #1} (#2) #3}
\def\PLA #1 #2 #3{{\sl Phys. Lett.} {\bf A #1} (#2) #3}
\def\PLB #1 #2 #3{{\sl Phys. Lett.} {\bf B #1} (#2) #3}
\def\JMP #1 #2 #3{{\sl J. Math. Phys.} {\bf #1} (#2) #3}
\def\PTP #1 #2 #3{{\sl Prog. Theor. Phys.} {\bf #1} (#2) #3}
\def\SPTP #1 #2 #3{{\sl Suppl. Prog. Theor. Phys.} {\bf #1} (#2) #3}
\def\AoP #1 #2 #3{{\sl Ann. of Phys.} {\bf #1} (#2) #3}
\def\PNAS #1 #2 #3{{\sl Proc. Natl. Acad. Sci. USA} {\bf #1} (#2) #3}
\def\RMP #1 #2 #3{{\sl Rev. Mod. Phys.} {\bf #1} (#2) #3}
\def\PR #1 #2 #3{{\sl Phys. Reports} {\bf #1} (#2) #3}
\def\AoM #1 #2 #3{{\sl Ann. of Math.} {\bf #1} (#2) #3}
\def\UMN #1 #2 #3{{\sl Usp. Mat. Nauk} {\bf #1} (#2) #3}
\def\FAP #1 #2 #3{{\sl Funkt. Anal. Prilozheniya} {\bf #1} (#2) #3}
\def\FAaIA #1 #2 #3{{\sl Functional Analysis and Its Application} {\bf
#1} (#2) #3}
\def\BAMS #1 #2 #3{{\sl Bull. Am. Math. Soc.} {\bf #1} (#2)
#3} \def\TAMS #1 #2 #3{{\sl Trans. Am. Math. Soc.} {\bf #1} (#2) #3}
\def\InvM #1 #2 #3{{\sl Invent. Math.} {\bf #1} (#2) #3}
\def\LMP #1 #2 #3{{\sl Letters in Math. Phys.} {\bf #1} (#2) #3}
\def\IJMPA #1 #2 #3{{\sl Int. J. Mod. Phys.} {\bf A #1} (#2) #3}
\def\AdM #1 #2 #3{{\sl Advances in Math.} {\bf #1} (#2) #3}
\def\RMaP #1 #2 #3{{\sl Reports on Math. Phys.} {\bf #1} (#2) #3}
\def\IJM #1 #2 #3{{\sl Ill. J. Math.} {\bf #1} (#2) #3}
\def\APP #1 #2 #3{{\sl Acta Phys. Polon.} {\bf #1} (#2) #3}
\def\TMP #1 #2 #3{{\sl Theor. Mat. Phys.} {\bf #1} (#2) #3}
\def\JPA #1 #2 #3{{\sl J. Physics} {\bf A#1} (#2) #3}
\def\JSM #1 #2 #3{{\sl J. Soviet Math.} {\bf #1} (#2) #3}
\def\MPLA #1 #2 #3{{\sl Mod. Phys. Lett.} {\bf A #1} (#2) #3}
\def\JETP #1 #2 #3{{\sl Sov. Phys. JETP} {\bf #1} (#2) #3}
\def\JETPL #1 #2 #3{{\sl  Sov. Phys. JETP Lett.} {\bf #1} (#2) #3}
\def\PHSA #1 #2 #3{{\sl Physica} {\bf A #1} (#2) #3}
\def\CQG #1 #2 #3{{\sl Class. Quantum Grav.} {\bf #1} (#2) #3}
\def\SJNP #1 #2 #3{{\sl Sov. J. Nucl. Phys. (Yadern.Fiz.)} 
{\bf #1} (#2) #3}
\def\bea{\begin{eqnarray}}
\def\ena{\end{eqnarray}}
\def\a{\alpha}\def\b{\beta}\def\g{\gamma}\def\d{\delta}\def\e{\epsilon}

\def\l{\lambda}\def\L{\Lambda}
\def\th{\theta}
\newcommand{\p}[1]{(\ref{#1})}
\begin{document}
\renewcommand{\thefootnote}{\fnsymbol{footnote}}
\thispagestyle{empty}
\begin{flushright}
Preprint DFPD 96/TH/49\\
hep-th/96xxxxx\\
\end{flushright}

\medskip
\begin{center}
{\large\bf An $n=(1,1)$ super--Toda Model Based on $OSp(1|4)$}\\
\vspace{0.3cm}

{\bf Dmitrij Sorokin\footnote{On leave from Kharkov Institute of
Physics and Technology, Kharkov, 310108, Ukraine.}
and Francesco Toppan,}\\
\vspace{0.2cm}
{\it Dipartimento di Fisica ``Galileo Galilei'',\\
Universit\`a degli Studi di Padova\\
and $~^{(\star )}$ INFN, Sezione di Padova;\\
Via F. Marzolo 8, 35131 Padova, Italy.}\\
e--mail: sorokin@pd.infn.it\\
~~~~~~~~~~toppan@pd.infn.it\\ 

\vspace{0.5cm}
%\vspace{0.3cm}
{\bf Abstract}
\end{center}
We show that a Hamiltonian reduction of affine Lie superalgebras 
having bosonic simple roots (such as $OSp(1|4)$) {\it does} produce 
supersymmetric Toda models, with superconformal symmetry being 
nonlinearly realised for those fields of the Toda system which are 
related to the bosonic simple roots of the superalgebra. A fermionic
$b-c$ system of conformal spin $({3\over 2},-{1\over 2})$ is a natural 
ingredient of such models.

\newpage
\setcounter{page}1
\renewcommand{\thefootnote}{\arabic{footnote}}
\setcounter{footnote}0
\section{Introduction}
A systematic way of getting exactly--solvable Toda models \cite{ls} 
is to carry out a Hamiltonian reduction of affine Lie algebras  
or Wess--Zumino--Novikov--Witten models  
associated with them \cite{bal,f}. 
The Hamiltonian reduction consists in imposing 
(first--class) constraints on components of algebra--valued currents. 
This procedure is of particular importance for understanding the 
underlying algebraic structure of the  Toda models (such as 
$W$--algebra extension of $d=2$ conformal symmetry), for simplifying 
the construction of the general solution of the Toda equations and 
solving the quantization problem (see \cite{f} for a review).

The Hamiltonian reduction procedure is also applicable to the case of 
affine superalgebras and corresponding super--WZNW models \cite{drs}.
On one hand it serves, for instance, as a powerful method for studying 
superconformal and $W$--structure of superstring theory 
\cite{bersh,sev},
and on the other hand, as a result 
of the Hamiltonian reduction  one gets super--Toda models and 
super--$W$--algebras \cite{drs,hollo}. 
However, in contrast to the bosonic case not all 
superalgebras have been involved into the production of 
{\it supersymmetric}
Toda models yet. This is connected with a well--known fact that the 
standard Hamiltonian reduction (which implies imposing (first--class)
constraints directly on basic (super)currents of the affine 
(super)algebras) leads 
to Toda models with explicitly broken two--dimensional supersymmetry if 
the affine superalgebras subject to the Hamiltonian reduction contain 
not only odd (or fermionic) simple roots but also even (bosonic) simple 
roots in any simple root system of their generators \cite{hollo}. This also 
takes place in the case of the bosonic superaffine Lie algebras, i.e. 
algebras whose currents are superfields in a $d=2$ superspace  and 
whose root systems are always bosonic.

In \cite{wz} it was demonstrated how one can generalize the Hamiltonian 
reduction procedure to get super--Toda--like equations from super--WZNW 
models based on bosonic Lie groups. The simplest example of an 
$n=(1,1)$ supersymmetric $Sl(2,{\bf R})$ WZNW model was considered 
in detail, and supersymmetric version of the Liouville equation 
alternative to the standard one \cite{kulish} was obtained by 
imposing nonlinear constraints on $Sl(2,{\bf R})$ supercurrents. 
%The constraints 
%turned out to be a mixture of first-- and 
%second--class constraints. (Usually, when performing the 
%Hamiltonian reduction, one deals with linear first--class 
%constraints \cite{f,drs}). 
Such a Liouville system emerged before as a system of 
equations for a single bosonic (Liouville) and two fermionic physical 
degrees of freedom of a classical Green--Schwarz superstring propagating 
in $N=2$, $D=3$ flat target superspace \cite{lio} 
\footnote{One should not confuse this Liouville system describing classical 
physical modes of the string with (super)Liouville modes arising as 
anomalies of quantized noncritical strings \cite{pol}.}. Worldsheet $n=(1,1)$ 
superconformal symmetry of the system is nonlinearly realized on the 
bosonic Liouville field and the two fermionic fields, the latter 
transform as Goldstone fermions and have negative conformal spin 
$-{1\over 2}$ unusual for matter fields. 
This means that supersymmetry is broken spontaneously.
The corresponding 
stress--tensor of conformal spin 2 and the supersymmetry current of spin 
3/2 are constructed out of the spin 0 Liouville field, the spin 
$-{1\over 2}$ field and its conjugate momentum of spin ${3\over 2}$
(see \cite{wz} for the details). Such a so called fermionic $b-c$ system 
plays a particular role in superconformal theory and, for instance, in 
constructing a theory of embedding fermionic strings \cite{grs,bersh,ber}. 
Thus, the 
generalization of the Hamiltonian reduction in application to 
(super)affine (super)algebras with simple bosonic roots allows one to 
naturally obtain (already at the classical level and without untwisting 
ghosts \cite{ber}) superconformal $b-c$ systems which might correspond to 
unexplored vacua of string theory. In \cite{bersh}
an alternative reduction procedure was used to get $n=2$ superconformal 
$b-c$ structures and realizations of super--$W_n$ algebras from the 
$Sl(n|n-1)$ superalgebras containing simple 
bosonic roots in a Gauss decomposition.

In the present article we consider the Hamiltonian reduction of a 
current superalgebra in $n=(1,1)$, $d=2$ superspace
based on the rank 2 superalgebra $OSp(1|4)$, which is the simplest 
example of the superalgebras with simple bosonic roots (its simple root 
system consists of one bosonic and one fermionic simple root 
\cite{kac,sorba}).
Note that Hamiltonian reduction of affine $OSp(1|4)$ currents in $d=2$ 
bosonic space, which results in a nonsupersymmetric $W$--algebra 
structure, has been carried out in \cite{ms}. Below we shall show how, by 
imposing appropriate nonlinear constraints on $n=(1,1)$ supercurrents 
corresponding to the bosonic simple roots, one can circumvent the 
obstacle caused by the presence of the bosonic simple 
roots \cite{hollo} and 
get an $n=(1,1)$ supersymmetric Toda model with superconformal symmetry 
being nonlinearly realized for those components of the Toda system which 
are related to the bosonic simple root of $OSp(1|4)$. 

The set of fields of the super--Toda model consists of two bosonic fields 
of conformal spin 0, which are coupled to two fermionic fields of spin 
$1\over 2$ and two (left--right) chiral Goldstone fermions of spin 
-$1\over 2$. These fields are accompanied by a decoupled pair of free 
left--right chiral fermions of spin $3\over 2$. The spin $3\over 2$
and -$1\over 2$ fields form the fermionic $b-c$ system. In the absence 
of the fermions the bosonic fields obey the equations of motion of 
a Toda model based on $Sp(4)$. 

All the 
constraints imposed on $OSp(1|4)$ supercurrent components are of 
the first class 
except one fermionic second--class constraint. A complete gauge fixing 
of local symmetries corresponding to the first--class constraints 
results in a set of five two--dimensional holomorphic and antiholomorphic
current fields which survive the Hamiltonian reduction. These are (in 
the holomorphic sector) a conformal spin ${-{1\over 2}}$ field $c$, spin 
$3\over 2$ field $G$, spin 2 field $T$, spin $5\over 2$ field $W_{5\over 
2}$ and spin 4 field $W_4$. The currents $G$ and $T$ generate $n=1$ 
superconformal symmetry of the Toda model, while entire set of the 
currents might generate a super--$W$--algebra. We shall discuss this 
point in Conclusion upon establishing links (via truncation) of our 
$OSp(1|4)$ model with the Hamiltonian 
reduction of affine $OSp(1|4)$ currents in $d=2$ bosonic space \cite{ms},
with the supersymmetric versions of the Liouville model based on the 
$OSp(1|2)$ \cite{kulish} and $Sl(2,{\bf R})$ \cite{lio}, and with 
a bosonic Toda model based on $Sp(4)$ \cite{bal}.

\section{Affine $OSp(1|4)$ superalgebra}

$OSp(1|4)$ is a 14--dimensional superalgebra of rank 2. It has 10 
bosonic and 4 fermionic generators. Two bosonic generators $H_\a$ 
form a Cartan subalgebra 
(index $\a=1,2$ or denote $b$ and $f$ which stand, respectively, for 
``bosonic'' and ``fermionic''). The $OSp(1|4)$  root 
system consists of one positive (negative) fermionic simple root 
and one positive (negative) bosonic simple root. 
We shall denote   
fermionic and bosonic generators associated with the simple roots,
respectively, $E_{\pm f}$
and $E_{\pm b}$. The remaining 
generators (collectively denoted as $E_{+i}$, $E_{-i}$ ($i=1,2,3,4$)) 
correspond to higher positive and negative roots.

The 
(anti)commutation properties of the $OSp(1|4)$ generators  
can be derived by use of the fundamental
representation of $OSp(1|4)$ realized on $(4+1)\times(4+1)$ 
supermatrices (with $4$ bosonic and $1$ fermionic indices) 
$$
\begin{array}{ll}
H_1 = e_{11} - e_{22} - e_{33} + e_{44}; & H_2 = e_{33} - e_{44}\\
E_{+b} = e_{13} + e_{42}; & E_{-b} = e_{24} + e_{31}\\
E_{+f} = e_{35} + e_{54}; & E_{-f} = e_{53} - e_{45}
\end{array}
$$
$$
\begin{array}{ll}
E_{+1} = 2 e_{34}; & E_{-1} = -2 e_{43}\\
E_{+2} = e_{15} - e_{52}; & E_{-2} = - (e_{51} + e_{21})\\
E_{+3} = 2 (e_{14} - e_{32}); & E_{-3} = 2 (e_{41} - e_{23})\\
E_{+4} = - 4 e_{12}; & E_{-4} = 4 e_{21} 
\end{array}
$$
In the above formulas we denote with $e_{ij}$ the supermatrices having
as entries $c_{kl}= \delta_{ik}\delta_{jl}$. Notice that discarding
the fermionic generators $E_{\pm f}, E_{\pm 2}$ we get  the bosonic 
$sp(4)$ subalgebra of $OSp(1|4)$, which is then realised on bosonic 
$(4\times 4)$ matrices. 
For our 
purposes we shall mainly need the form of the $OSp(1|4)$ Cartan matrix
which is given by \cite{sorba}:
\begin{eqnarray}\label{cm}
K_{\a\b}=&
\left( \begin{array}{cc}
2 & -1\\
-1 & 1
\end{array}\right).
\end{eqnarray}

The holomorphic and antiholomorphic copies of the classical 
$n=(1,1)$, $d=2$ affine $OSp(1|4)$ superalgebra are generated 
by fermionic supercurrents
\begin{equation}\label{Psi}
\Psi(Z)=\Psi^{0\a}H_\a + \Psi^{-f}E_{-f}+\Psi^{-b}E_{-b}+\Psi^{+f}E_{+f}
+\Psi^{+b}E_{+b}+\sum_i\Psi^{-i}E_{-i}+\sum_i\Psi^{+i}E_{+i},
\end{equation}
$$
\bar\Psi(\bar Z)=\bar\Psi^{0\a}H_\a + \bar\Psi^{-f}E_{-f}+
\bar\Psi^{-b}E_{-b}+\bar\Psi^{+f}E_{+f}
+\bar\Psi^{+b}E_{+b}+\sum_i\bar\Psi^{-i}E_{-i}+\sum_i\bar\Psi^{+i}E_{+i}.
$$
$n=(1,1)$, $d=2$ superspace is parametrized by supercoordinates
$Z=(z,\th)$, $\bar Z=(\bar z, \bar\th)$
(where $z,~\bar z$ are bosonic and $\th,~\bar\th$ are fermionic 
light--cone coordinates).  
The supercovariant derivatives $D={\partial\over{\partial\th}}+
i\th{\partial\over{\partial z}},~ \partial={\partial\over{\partial z}}$ 
and $\bar D={\partial\over{\partial\bar\th}}+i\bar\th
{\partial\over{\partial\bar 
z}},~\bar\partial={\partial\over{\partial{\bar z}}}$ 
form the following algebra
\begin{equation}\label{D}
\{D,D\}=2i{\partial\over{\partial z}},\qquad 
\{\bar D,\bar D\}=2i{\partial\over{\partial\bar z}}, \qquad
\{D,\bar D\}=0.
\end{equation}
The $n=(1,1)$ superconformal transformations 
of $Z, \bar Z$ and their derivatives are:
$$
Z'=Z'(Z),~~~~~{\bar Z}'={\bar 
Z}'(\bar Z);
$$ 
\begin{equation}\label{sc}
D'=e^{-\L}D,\qquad {\bar D}'=e^{-\bar\L}\bar D, \qquad 
{\rm where}~~~~\L=\log D\th'(Z),~~~~\bar\L=\log\bar D{\bar\th}'(\bar Z);
\end{equation}
and
$$
\Psi'(Z')=e^{-\L}\Psi(Z), \qquad \Psi'(Z')=e^{-\bar\L}\bar\Psi(\bar Z).
$$
Under (left--right) affine $OSp(1|4)$ transformations the 
supercurrents transform as follows:
\begin{equation}\label{sa}
\Psi'(Z)=g^{-1}_l(Z)\Psi g_L(Z)+ig^{-1}_LDg_L;
\qquad \bar\Psi'(Z)=g_R(\bar Z)\bar\Psi g^{-1}_R(\bar Z)+
ig^{-1}_R\bar Dg_R.
\end{equation}
Poisson brackets of the supercurrents are \cite{drs}
\begin{equation}\label{pbs}
\left\{str(A\Psi(X)),str(B\Psi(Y))\right\}_{PB}
=-\d(X-Y)str[A,B]\Psi(Y)-iD_X\d(X-Y)str(AB),
\end{equation}
where $A$ and $B$ stand for the $OSp(1|4)$ generators; 
$X=(z_1,\th_1)$, $Y=(z_2,\th_2)$, \\
$\d(X-Y)=\d(z_1-z_2)(\th_1-\th_2)$, and
$D_X={\partial\over\partial\th_1}+i\th_1{\partial\over\partial z_1}$.
It is implied that the Poisson brackets are equal--time, i.e.
$z_1+\bar z_1=z_2+\bar z_2$. 
(The Poisson brackets of $\bar\Psi(\bar Z)$ have the same form).
 
By definition the affine $OSp(1|4)$ supercurrents satisfy the (anti)chirality 
conditions, which are preserved under the superconformal \p{sc} and 
superaffine \p{sa} transformations:
\begin{equation}\label{chi}
\bar D\Psi(Z)=0, \qquad D\bar\Psi(\bar Z)=0.
\end{equation}

We assume these conditions to arise as the equations of motion of an 
$OSp(1|4)$ WZNW model so that $\Psi(Z)$ and $\bar\Psi(\bar Z)$
are expressed in terms of an $OSp(1|4)$ supergroup element $G(Z,\bar 
Z)$:
\begin{equation}\label{gp}
\Psi=-iDGG^{-1}, \qquad \bar\Psi=iG^{-1}\bar DG.
\end{equation}

To get super--Toda equations from \p{chi} one should express the 
supercurrents in terms of superfields of the Gauss decomposition of 
$G(Z,\bar Z)$:
\begin{equation}\label{gd}
G=e^{\b^\a E_{+\a}+\b^i E_{+i}}e^{\Phi_\a H_\a}e^{\g^\a E_{-\a}+
\g^i E_{-i}}
\end{equation}
and perform a Hamiltonian reduction, i.e. to impose constraints on 
$OSp(1|4)$--valued components of the supercurrents. Note that
they can be imposed at a purely algebraic level, i.e. 
without referring to the equation (\ref{gp}).

Since the form of $\Psi(Z)$ and $\bar\Psi(\bar Z)$ components in terms of 
superfields $\b(Z,\bar Z)$, $\g(Z,\bar Z)$ and $\Phi(Z,\bar Z)$ is 
essentially simplified upon imposing the constraints, we shall 
first discuss these constraints and then write down expressions for
 $\Psi(Z)$ and $\bar\Psi(\bar Z)$ in terms of the Gauss decomposition 
superfields which, being substituted into \p{chi}, give rise to the 
super--Toda equations.

The Hamiltonian reduction procedure \cite{bal,f,drs} prescribes
constraining holomorphic supercurrents associated with simple negative 
roots to be nonzero constants, and the ones associated with the negative
nonsimple roots are put equal to zero. (In the antiholomorphic sector 
the positive--root supercurrent components are constrained).

In the case of superalgebras \cite{hollo,drs}
the Hamiltonian reduction causes no problems 
with preserving supersymmetry if a root system of the superalgebra 
contains only simple fermionic roots. Then the corresponding 
supercurrents (such as $\Psi^{-f}$ in \p{Psi}) are bosonic and one can 
put them equal to constants without violating supersymmetry. But if, as 
in the case of $OSp(1|4)$, a root system of a superalgebra contains 
bosonic simple roots, the corresponding supercurrents (such as 
$\Psi^{-b}$) are fermionic and constraining them to be Grassmann 
constants or equal to the Grassmann coordinate $\th$ \cite{hollo}
one explicitly breaks supersymmetry of the model.

In \cite{wz} it was proposed to overcome this problem by constructing 
$OSp(1|4)$ bosonic supercurrents out of the fermionic supercurrents as 
prescribed by the Maurer--Cartan equations (see \cite{wz} for details):
\begin{equation}\label{J}
J(Z)=D\Psi-i\Psi\Psi, \qquad
\bar J(\bar Z)=\bar D\bar\Psi -i\bar\Psi\bar\Psi
\label{jjj}
\end{equation}
(where matrix multiplication is implied) and then put equal to 
constants the superalgebra--valued components of $J$ and $\bar J$
corresponding, respectively, to the negative and positive bosonic 
simple--root generators $E_{-b}$ and $E_{+b}$.

Since the constraints imposed this way have a superfield form, 
supersymmetry is not explicitly broken but, as we shall see, it is 
nonlinearly realized on part of the fields which survive the Hamiltonian 
reduction.

Thus we impose on fermionic supercurrents \p{chi} the following set of 
constraints:
\begin{equation}\label{c}
\Psi^{-f}=\mu_f, \qquad \Psi^{-i}=0, ~~~~ J^{-b}\equiv 
D\Psi^{-b}+2i\Psi^{0b}\Psi^{-b}=\mu_b;
\end{equation}
$$
\bar\Psi^{+f}=\nu_f, \qquad \bar\Psi^{+i}=0, ~~~~ \bar J^{+b}\equiv 
\bar D\bar\Psi^{+b}-2i\bar\Psi^{0b}\bar\Psi^{+b}=\nu_b,
$$ 
where the factor 2 comes from the Cartan matrix \p{cm}, and $\mu_\a$ and
$\nu_\a$ ($\a=b,f$ or (1,2)) are arbitrary constants.

We postpone the Hamiltonian analysis of these constraints to Section 4 
and turn to getting the supersymmetric Toda system associated with 
$OSp(1|4)$.

\section{$OSp(1|4)$ super--Toda equations.}

To derive the $n=(1,1)$ supersymmetric Toda equations from \p{chi} and 
\p{c} one should write down supercurrents 
$\Psi^{0\a}=(\Psi^{0f},\Psi^{0b})$, $\Psi^{-\a}=(\Psi^{-f},\Psi^{-b})$
and their antiholomorphic counterparts in terms of fields of the Gauss 
decomposition \p{gd}. With taking into account the constraints we get
$$
\Psi^{0f}=-iD\Phi_2+\mu_f\b^{f},\qquad \Psi^{0b}=
-iD\Phi_1+\b^{b}\Psi^{-b}, 
$$
$$
\Psi^{-b}=-iD\g^be^{-K_{b\a}\Phi^\a},
\qquad \Psi^{-f}=-iD\g^fe^{-K_{f\a}\Phi^\a}=\mu_f;
$$
$$
\bar\Psi^{0f}=i\bar D\Phi_2+\nu_f\g^{f},\qquad \bar\Psi^{0b}=
iD\Phi_1+\g^{b}\bar\Psi^{+b}, 
$$
\begin{equation}\label{gau}
\bar\Psi^{+b}=i\bar D\b^be^{-K_{b\a}\Phi^\a}, \qquad 
\bar\Psi^{+f}=i\bar D\b^fe^{-K_{f\a}\Phi^\a}=\nu_f.
\end{equation}
In \p{gau} we assume summation over $\a=1,2$, and
remember that $b$ and $f$ correspond, respectively, 
to the single bosonic and fermionic 
simple root and stand for the index 1 and 2 of the Cartan matrix 
components \p{cm}.

Hitting the r.h.s of $\Psi^{0f}$ and $\Psi^{0b}$ in \p{gau} 
with the supercovariant derivative $\bar D$ and taking into account the 
chirality conditions \p{chi}, the form of the Cartan matrix \p{cm} 
and other expressions in \p{gau} we get the 
following system of superfield equations:\footnote{It is possible,
and indeed preferable, to treat $\Psi^{-b}$, $\bar\Psi^{+b}$ as 
independent superfields; in this case $\gamma^b$, $\beta^b$ do not enter 
the equations (\ref{equ1},\ref{equ2}).} 
\begin{equation}\label{st}
\bar DD\Phi_2=i\mu_f\nu_fe^{\Phi_2-\Phi_1},
\qquad
\bar DD\Phi_1=e^{2\Phi_1-\Phi_2}\bar\Psi^{+b}\Psi^{-b},\qquad 
\bar D\Psi^{-b}=0=D\bar\Psi^{+b}.
\label{equ1}
\end{equation}
To Eqs. \p{st} one must add the constraints on $J^{-b}$ and $\bar 
J^{+b}$ in \p{c}, which now take the form:
\begin{equation}\label{cn}
D\Psi^{-b}+2D\Phi_1\Psi^{-b}=\mu_b, \qquad
\bar D\bar\Psi^{+b}+2 
\bar D\Phi_1\bar\Psi^{+b}=\nu_b.
\label{equ2}
\end{equation}
The system of equations \p{st} and \p{cn} is the $OSp(1|4)$ 
super--Toda system we have been looking for.

Let us relate the super--Toda system \p{st}, \p{cn} to some known 
(super)--Toda equations.

The $OSp(1|4)$--based Toda system with explicitly broken $n=(1,1)$ 
supersymmetry \cite{hollo} is recovered by putting $\Psi_{-b}=\th$ and 
$\bar\Psi_{+b}=\bar\th$. Notice that it contains less fermionic degrees 
of freedom than in \p{st} and \p{cn}. 
If we in addition put equal to zero all higher 
superfield components of $\Phi_1$ and $\Phi_2$ (except the leading 
ones), we get a bosonic Sp(4) Toda model \cite{bal}.

The $n=(1,1)$ super--Liouville equation \cite{kulish} based on an 
$OSp(1|2)$ subgroup of $OSp(1|4)$ is obtained as a truncation of the 
first equation of \p{st} by putting in \p{st} and \p{cn} 
$\Psi^{-b}=\bar\Psi^{+b}=\mu_b=\mu_b=0$ and $\Phi_1=constant$.

The alternative version \cite{lio,wz} of the $n=(1,1)$ super--Liouville 
system based on an $Sl(2,R)$ subgroup of $OSp(1|4)$ arises as a 
truncated form of Eqs. \p{st}, \p{cn} where $\Phi_2=constant$ and 
$\mu_f=\nu_f=0$. (Note that putting some of $\Phi_\a$ equal to constants 
and $\mu_\a=\nu_\a=0$ is equivalent to putting to zero corresponding 
supercurrents of $OSp(1|4)$,  which thus truncates the affine $OSp(1|4)$ 
down to its subalgebras).

Now let us discuss in more detail properties of Eqs. \p{st}, \p{cn}.
The system is invariant under the following $n=(1,1)$ superconformal 
transformations \p{sc} of the superfields:
$$
\Psi'^{-b}(Z')=e^\L\Psi^{-b},\qquad 
\Psi'^{+b}(Z')=e^{\bar\L}\bar\Psi^{+b},
$$
$$
\Phi'_1(Z')=\Phi_1-3(\L+\bar\L)-
{5\over 2}(D\L\Psi^{-b}+\bar D\bar\L\bar\Psi^{+b}),
$$
\begin{equation}\label{msc}
\Phi'_2(Z')=\Phi_2-4(\L+\bar\L)-
{5\over 2}(D\L\Psi^{-b}+\bar D\bar\L\bar\Psi^{+b}).
\end{equation}
 
One can see that the superconformal properties of the supercurrents
$\Psi$, $\bar\Psi$ have changed. This is a consequence of imposing the 
constraints \p{c} which break part of affine $OSp(1|4)$ symmetries 
\cite{bal,f,drs}. For instance the superaffine transformations 
corresponding to the $OSp(1|4)$ Cartan subalgebra are combined with 
initial superconformal transformations of the supercurrents to produce 
the modified superconformal transformations \p{msc} which preserve the 
form of the constraints \p{c}. The remaining local gauge symmetries of 
the model are generated by those constraints in \p{c} which are of the 
first class. We discuss this point in the next Section.

Eqs. \p{msc} tell us that superconformal symmetry is nonlinearly 
realized on superfields of the super--Toda model.  
In order that one of the Cartan 
superfields transforms independently of $\Psi^{-b},{\bar\Psi}^{+b}$,
it is convenient to redefine the fields as follows:
$$
{\hat\Phi}_1={\Phi_1\over 3}
-{2\over 3}(D\Phi_1\Psi^{-b}+\bar D\Phi_1\bar\Psi^{+b}),
\qquad
{\hat\Phi}'_1(Z')={\hat\Phi}_1-\L-\bar\L-
{1\over 2}(D\L\Psi^{-b}+\bar D\bar\L\bar\Psi^{+b});
$$
\begin{equation}\label{fr}
{\hat\Phi}_2=\Phi_2-\Phi_1, \qquad
{\hat\Phi}'_2(Z')={\hat\Phi}_2-\L-\bar\L,
\end{equation}
where on the r.h.s. of \p{fr} the superconformal transformations of the 
redefined superfields are reproduced. In terms of ${\hat\Phi}_\a$ Eqs. \p{st},
\p{cn} take the form \footnote{In what follows we put $\nu_\a=\mu_\a=1$ 
without loosing generality}:
$$
\bar DD{\hat\Phi}_2=ie^{{\hat\Phi}_2}
(1+ie^{3{\hat\Phi}_1-2{\hat\Phi}_2}\bar\Psi^{+b}\Psi^{-b}),
$$
\begin{equation}\label{stm}
\bar DD{\hat\Phi}_1={1\over 3}e^{3{\hat\Phi}_1-{\hat\Phi}_2}
\bar\Psi^{+b}\Psi^{-b},\qquad 
\bar D\Psi^{-b}=0=D\bar\Psi^{+b}.
\end{equation}
\begin{equation}\label{cnm}
D\Psi^{-b}+2D{\hat\Phi}_1\Psi^{-b}=1, \qquad
\bar D\bar\Psi^{+b}+2\bar D{\hat\Phi}_1\bar\Psi^{+b}=1.
\end{equation}

The constraints \p{cnm} are explicitly solvable \cite{lio,wz} and 
result in the following form of $\Psi^{-b}$, $\bar\Psi^{+b}$ and
${\hat\Phi}_1$:
$$
\Psi^{-b}=c(z)+\th(1+ic\partial c), \qquad
\bar\Psi^{+b}=\bar c(\bar z)+
\bar\th(1+i\bar c\bar\partial\bar c),
$$
\begin{equation}\label{com}
{\hat\Phi}_1={\hat\phi}_1(z,\bar z)+{i\over 2}\th 
e^{-2{\hat\phi}_1}\partial(e^{2{\hat\phi}_1}c)+
{i\over 2}\bar\th e^{-2{\hat\phi}_1}\bar\partial(e^{2{\hat\phi}_1}\bar c)+
\th\bar\th c\bar c\partial\bar\partial{\hat\phi}_1.
\end{equation}
Note that the superconformal symmetry is spontaneously
broken since under the superconformal transformations \p{sc}, \p{msc}
(with $z~\to~z-\l(z)-i\th\e$, $\bar z~\to~\bar z-\bar\l(\bar 
z)-i\bar\th\bar\e$,
$\th\rightarrow\th-\e(z)-\th\partial\l$, 
$\bar\th\rightarrow\bar\th-\bar\e(\bar 
z)-\bar\th\bar\partial\bar\l$ and 
$\L(Z)=-\partial\l-i\th\partial\e)$ the spinor 
fields $c(z)$ and $\bar c(z)$ transform as Goldstone fermions of 
conformal spin $-{1\over 2}$:
\begin{equation}\label{sconf}
\d_\l c=\l\partial c -{1\over 2}c\partial\l, \qquad
\d_\e c=c(z)+\e(z)+i\e(z)c\partial c,
\end{equation}
$$
\d_{\bar\l}\bar c=\bar\l\bar\partial\bar c 
-{1\over 2}\bar c\bar\partial\bar\l,
\qquad
\d_{\bar\e}\bar c=\bar c(\bar z)+\bar\e(\bar z)+
i\bar\e(\bar z)\bar c\partial\bar c.
$$

Finally we present the component form of Eqs. \p{stm},  where 
$\phi_2={\hat\Phi}_2|_{\th,\bar\th=0}$, 
$\psi=iD{\hat\Phi}_2|_{\th,\bar\th=0}$,
$\bar\psi=i\bar D{\hat\Phi}_2|_{\th,\bar\th=0}$ and ${\phi}_1=3{\hat\phi}_1-
ic({1\over 2}\partial c+\psi)-
i\bar c({1\over 2}\bar\partial\bar c+\bar\psi)$ (this $\phi_1$ should 
not be confused with the leading component of $\Phi_1$ in \p{st}):

%%%%%%%%%%%Toda equations without redefinition of $\hat\phi_1$%%%%%
%%%%%%%%%%% not included into the article %%%%%%%%%%%%%%%%%%%%%%%%%%
%$$
%\bar\partial\partial{\hat\phi}_1=-{1\over 3}e^{3{\hat\phi}_1-\phi_2
%-ic({1\over 2}\partial c+\psi)-
%i\bar c({1\over 2}\bar\partial\bar c+\bar\psi)}
%-{i\over 3}e^{3{\hat\phi}_1}\bar cc,
%$$
%$$
%\bar\partial\partial\phi_2=
%-e^{\phi_2}(e^{\phi_2}+i\bar\psi\psi)
%+e^{3{\hat\phi}_1-\phi_2-
%ic({1\over 2}\partial c+\psi)-
%i\bar c({1\over 2}\bar\partial\bar c+\bar\psi)},
%$$
%\begin{equation}\label{comp}
%\bar\partial\psi=e^{\phi_2}\bar\psi-e^{3{\hat\phi}_1-\phi_2}c(1-i\bar c
%({1\over 2}\bar\partial \bar c+\bar\psi)), \qquad
%\partial\bar\psi=-e^{\phi_2}\psi-e^{3{\hat\phi}_1-\phi_2}\bar c(1-ic
%({1\over 2}\partial c+\psi)),
%\end{equation}
%%%%%%%%%%%%%%%%%%%%%%%%%%%%%%%%%%%%%%%
$$
\bar\partial\partial{\phi}_1=-e^{{\phi}_1-\phi_2}
-ie^{{\phi}_1}\bar cc -
i\partial(e^{\phi_2}c\bar\psi)+i\bar\partial(e^{\phi_2}\bar 
c\psi),
$$
$$
\bar\partial\partial\phi_2=
-e^{\phi_2}(e^{\phi_2}+i\bar\psi\psi)
+e^{{\phi}_1-\phi_2},
$$
\begin{equation}\label{comp}
\bar\partial\psi=e^{\phi_2}\bar\psi-e^{{\phi}_1-\phi_2}c, \qquad
\partial\bar\psi=-e^{\phi_2}\psi-e^{{\phi}_1-\phi_2}\bar c, \qquad
\bar\partial c=0,\qquad \partial\bar c=0.
\end{equation}
From Eqs. \p{comp} it follows that the free Goldstone fermions 
$c(z)$, $\bar c(\bar z)$ contribute to the r.h.s. of equations of 
motion of $\phi_1$, $\psi$ and $\bar\psi$. This is in contrast to the 
simplest case of the Hamiltonian reduction of this kind applied to the 
affine superalgebra $sl(2,{\bf R})$, which leads to the super--Liouville
system with completely decoupled bosonic and fermionic sector \cite{wz}.

\section{Constraints and symmetries of the model}

Let us now turn to the consideration of the Hamiltonian properties of 
the constraints \p{c}, the resulting number of independent fields of the 
model and their symmetry properties.

One can directly check that all the constraints \p{c} commute with each 
other with respect to the Poisson brackets \p{pbs} except for the Grassmann 
component of $J^{-b}-\mu_b=0$ (and $\bar J^{+b}-\nu_b=0$ in the 
antiholomorphic sector) whose Poisson bracket with itself produces the $
\d$--function on the r.h.s. of \p{pbs}. Thus all the constraints \p{c} 
are of the first class except this fermionic second--class constraint. 
Analogous situation encountered in the $Sl(2,{\bf R})$ case has been 
described in detail in \cite{wz}. 

The first--class constraints generate local gauge transformations of the 
$OSp(1|4)$ supercurrent components that have not been constrained. This 
allows one to impose on some of these currents gauge  
conditions. The number of the gauge conditions is equal 
to the number of the first--class constraints. This reduces the initial 
number of the supercurrent components to an independent set of bosonic 
and fermionic current fields. In the case under consideration we 
initially have $14\times 2=28$ bosonic and fermionic fields. The 
$5\times 2=10$ first--class constraints $(\Psi^{-i}=0, (i=1,...,4),
\Psi^{-f}=\mu_f)$ together with corresponding gauge conditions eliminate
$20=10\times 2$ degrees of freedom, while the constraint $J^{-b}=\mu_b$
(which is a mixture of the first and second class constraint) eliminates 
2 bosonic and 1 fermionic degree of freedom.

One can gauge fix to zero those $OSp(1|4)$--valued supercurrents whose 
Poisson brackets \cite{bal,f,drs} with one of the first--class 
constraints contains a $\d$--function term. This indicates that 
under the corresponding gauge transformation the supercurrent undergoes 
an arbitrary shift proportional to the parameter of the transformation
and hence can be eliminated. 
This way one can put to zero the following components of \p{Psi} and 
\p{J} (in the holomorphic sector):
$$
\Psi^{0f}=\Psi^{+f}=\Psi^{+b}=J^{0b}|_{\th=0}=J^{+2}=J^{+3}=0.
$$
Note that despite of the non--manifestly supersymmetric form  the
constraint $J^{0b}|_{\th=0}=0$ is superconformal invariant (see 
\cite{wz}). 

Thus, we remain with $28-23=5$ independent currents both 
in the holomorphic and in the antiholomorphic sector of the 
constrained model. Among them we have 2 fermionic 
degrees of freedom corresponding to fields entering the 
fermionic supercurrents $\Psi^{+1}, \bar\Psi^{-1}$ (associated with the
$OSp(1|4)$ non--simple roots $E_{\pm1}$).
These two chiral and antichiral fermionic fields 
(which we call $b(z)$ and $\bar b(\bar z)$) 
have conformal spin ${3\over 2}$. By an appropriate change of variables 
they can be made canonical 
conjugate to the spin ($-{1\over 2}$) fields $c(z),\bar c(\bar z)$ 
with respect to Dirac brackets which one can construct by use of 
the constraints and the gauge conditions (see, for instance \cite{wz}).

Since $\Psi^{+1}$ and  $\bar\Psi^{-1}$ do not enter the 
super--Toda equations (\ref{equ1},~\ref{equ2}), 
the super--Toda system obtained in the previous Section 
does not contain all fields of the initial WZNW model which survive the 
Hamiltonian reduction. It is a closed superconformal subsystem of 
interacting fields which is part of a larger system which also
includes a completely decoupled pair of free chiral fermions of 
conformal spin ${3\over 2}$. This is a consequence of the presence in 
\p{c} of the second--class constraints, and is in contrast to the 
standard Hamiltonian reduction of bosonic Lie algebras and superalgebras 
with entirely fermionic simple root structure where the Cartan 
subalgebra fields of corresponding (super)--Toda model exhaust the number 
of remaining physical degrees of freedom \cite{bal,f,drs}.

In the holomorphic (as well as in the antiholomorphic) sector the 
above system 
of constraints and gauge fixing conditions is explicitly solved 
in terms of five unconstrained fields having conformal spins
$$
 (-{1\over 2},~ {3\over 2},~ 2,~ {5\over 2},~ 4 ).
$$
Let us call them
\bea\label{23}
(c, G, T, W_{5\over 2}, W_4). &&
\ena

On the one hand these fields are a combination of $OSp(1|4)$ 
supercurrent components and on the other hand they are expressed in 
terms of $\hat\phi_1,$ $\phi_2$, $\psi$, $b,$ $c$ and their derivatives.

The conformal spin 2 current $T(z)$ and the conformal spin ${3\over 2}$ 
current $G(Z)$ are given by:
$$
T=T_1+T_2-{{3i}\over 2}b\partial c-{i\over 2}\partial bc, 
$$
$$
T_1=k\left((\partial{\hat\phi}_1)^2-\partial^2{\hat\phi}_1\right), \qquad
T_2=k\left((\partial{\phi}_2)^2-\partial^2{\phi}_2+
{i\over 2}\psi\partial\psi\right);
$$
\begin{equation}\label{scr}
G=G_1+G_2;
\end{equation}
$$
G_1=ib+icT_1-bc\partial c - {k\over 4}c\partial c{\partial}^2c
-ik{\partial}^2c, \qquad
G_2=ik(\psi\partial\phi_2-i\partial\psi).
$$
They generate on the Dirac brackets the classical n=1 super--Virasoro 
algebra with central charge $c_{tot}=12 k$, where $k$ is a level 
\cite{bal,f,drs}
of affine $OSp(1|4)$ and 
 $c_{tot}$ is the sum of the two central charges $c_1$, $c_2$ 
related to the two 
independent super--Virasoro realizations ($T_1,G_1$) and ($T_2,G_2$).

The realization \p{scr} of the superconformal algebra involves the 
$(b-c)$ system which plays essential role in various physical
applications of superconformal theory (see, for example, 
\cite{bersh,grs,ber,wz}).

The fields $W_{5\over 2}$ and $W_4$ in \p{23} are primary fields with 
conformal spin ${5\over 2}$ and $4$, respectively. The derivation of 
their form in terms of the super--Toda fields and $b(z)$ is pretty 
cumbersome and we have not carried out it explicitly. 
Thus we have not yet 
checked that the full set of the fields \p{23} generate on the Dirac 
brackets a closed classical super--$\cal W$ algebra, though the 
consistency of the Hamiltonian reduction procedure performed and the 
relation of our model with that based on $OSp(1|2)$, $Sp(4)$ and 
$SL(2,{\bf R})$ group (as discussed in section 3) suggests possible 
existence of such an algebra. If this is indeed the case, such a new 
super--$\cal W$ algebra should be a supersymmetric extension of the 
so--called $WB_2$ algebra \cite{bks} generated by a stress 
energy tensor and two primary fields (analogous to $W_{5\over 2}$ and 
$W_4$) of conformal spin ${5\over 2}$ and $4$.
Indeed, in \cite{ms} the $WB_2$ algebra was obtained by imposing  
constraints on affine $OSp(1|4)$ currents which were taken to be 
ordinary two--dimensional fields and not superfields as in our case.
For that reason the Hamiltonian reduction of affine $OSp(1|4)$ carried out 
in \cite{ms} resulted in a non--supersymmetric $\cal W$--structure, and 
its supersymmetric extension might be generated by \p{23}.

In conclusion we have obtained the supersymmetric Toda model 
with peculiar symmetry properties
by applying 
the generalized Hamiltonian reduction procedure to the affine $OSp(1|4)$ 
superalgebra. 
%The $n=1$ superconformal symmetry is nonlinearly realized 
%on part of the super--Toda fields corresponding to the simple bosonic 
%root of $OSp(1|4)$ and is generated by the stress--energy tensor T and 
%the SUSY generator G \p{scr} containing ghost--like fermionic 
%$(b-c)$ fields of conformal spin ${3\over 2}$ and $({{-1}\over 2}$.
%Five current fields \p{23}, which survive the Hamiltonian reduction,
%are likely to generate a classical super--${\cal W}$ algebra containing 
%$WB_2$ as a subalgebra. 
From the physical point of view it seems of 
interest to study, within the frame of a geometrical approach 
\cite{bpstv,lio,wz,bs}, 
whether the model considered herein describes the dynamics of a superstring 
propagating in $D=4$ anti--De--Sitter superspace whose symmetry group is
$OSp(1|4)$.

\smallskip
\noindent
{\bf Acknowledgements}. The authors would like to thank E. Ivanov, 
S. Krivonos, Z. Popovicz, A. Sorin and M. Vasiliev for interest to this 
work and useful discussion. Work of D.S. was partially support by the 
INTAS grant N 93--493.


\begin{thebibliography}{99}
\bibitem{ls}
A. N. Leznov and M. V. Saveliev, \LMP 3 179 489; \CMP 74 1980 111.
\bibitem{bal}
P. Forg\'acs, A. Wipf, J. Balog, L. Feh\'er and L. O'Raifeartaigh,
\PLB 227 1989 214.\\
J. Balog, L. Feh\'er, L. O'Raifeartaigh, P. Forg\'acs and A. Wipf, 
\AoP 203 1990 76.
\bibitem{f}
L. Feh\'er, L. O'Raifeartaigh, P. Ruelle, I. Tsutsui and A. Wipf, 
\PR 222 1992 1.
\bibitem{drs}
F. Delduc, E. Ragoucy and P. Sorba, {\sl Commun. Math. Phys.} {\bf 146}
(1992) 403.
\bibitem{bersh}
M. Bershadsky, W. Lerche, D. Nemeschansky and N. P. Warner, \NPB 401 
1993 304.
\bibitem{sev}
E. Ragoucy, A. Sevrin and P. Sorba, Preprint ENSLAPP-555/95, VUB-TH.395, 
hep-th/9511049. 
\bibitem{hollo}
M. A. Olshanetsky, {\sl Commun. Math. Phys.} {\bf 88} (1983) 63.\\
J. Evans and T. Hollowood, \NPB 352 1991 723.
\bibitem{wz}
D. Sorokin and F. Toppan, Hamiltonian Reduction of Supersymmetric 
WZNW Models on Bosonic Groups and Superstrings, Preprint DFPD 96/TH/15,
Padua, March 1996, hep-th/9603187. (To appear in Nucl. Phys. {\bf B480}
n. 1,2).
\bibitem{kulish}
M. Chaichian and P. P. Kulish, \PLB 78 1978 413.
\bibitem{lio}
I. Bandos, D. Sorokin and  D. Volkov, {\sl Phys. Lett.} {\bf B372} 
(1996) 77. 
\bibitem{pol}
A. M. Polyakov, \PLB 103 1981 210.
\bibitem{grs}
B. Gato--Rivera and A. M. Semikhatov, \PLB 293 1992 72.
\bibitem{ber}
N. Berkovits and C. Vafa, \MPLA 9 1993 653.\\
N. Berkovits and N. Ohta, \PLB 334 1994 72.\\
F. Bastianelli, \PLB 322 1994 340.\\
F. Bastianelli, N. Ohta and J. L. Peterson, \PLB 327 1994 35.\\
N. Berkovits and C. Vafa, \NPB 433 1995 123.
\bibitem{kac}
V. G. Kac, {\sl Adv. Math.} {\bf 30} (1978) 85.
\bibitem{sorba}
L. Frappat, A. Sciarrino and P. Sorba, {\sl Comm. Math. Pyhs.} {\bf 121}
(1989) 457.
\bibitem{ms} 
J.O. Madsen and E. Ragoucy, ``Linearization of ${\cal W}$ algebras and
${\cal W}$ superalgebras, Preprint ENSLAPP-A-520/95, hep-th/9510061.
\bibitem{bks}
J.M. Figueroa-O'Farrill, S. Schrans and K. Thielemans, \PLB 263 1991 
378; S. Bellucci, S. Krivonos and A. Sorin, \PLB 347 1995 260. 
\bibitem{bpstv}
I.Bandos, P.Pasti, D.Sorokin, M.Tonin and D.Volkov,
{\sl Nucl. Phys.} {\bf B446} (1995) 79.\\
P. S. Howe and E. Sezgin, Preprint CERN-TH/96-200, 
hep-th/9607227.
\bibitem{bs}
I. Bakas and K. Sfetsos, \PRD 54 1996 3995.
\end{thebibliography}
\end{document}